\documentclass[conferences]{IEEEtran}
\usepackage{floatflt}
\usepackage{color}
\usepackage{cite}
\usepackage{graphicx}
\usepackage{amsmath}
\usepackage{amsthm}
\usepackage{amssymb}
\usepackage{verbatim}
\usepackage{psfrag,array}
\usepackage{subfigure}
\usepackage{multirow}
\usepackage{hhline}
\usepackage{stfloats}
\usepackage{amsmath}
\usepackage{epstopdf}
\usepackage{algorithmic}

\usepackage{algorithm}
\usepackage{algorithmic}

\newcommand{\p}[1]{\mathop{\mbox{\it p} } }

\renewcommand{\vec}[1]{\ensuremath{\boldsymbol{#1}}}
\newcommand{\be}{\begin{equation}}
\newcommand{\ee}{\end{equation}}
\newcommand{\ba}{\begin{array}}
\newcommand{\ea}{\end{array}}
\newcommand{\bea}{\begin{eqnarray}}
\newcommand{\eea}{\end{eqnarray}}
\newcommand{\bean}{\begin{eqnarray*}}
\newcommand{\eean}{\end{eqnarray*}}

\newcommand{\rmh}{^{\dag}}

\renewcommand{\Re}{\mathcal{R}}

\definecolor{white}{rgb}{1,1,1}

\newtheorem{theorem}{Theorem}

\begin{document}

\title{Sequential Channel Estimation in the Presence of Random Phase Noise in NB-IoT Systems}

\author
{
Fredrik Rusek and Sha Hu\\
Department of Electrical and Information Technology,
Lund University, Lund, Sweden\\
\{firstname.lastname\}@eit.lth.se
\vspace{-2mm}
}

\maketitle

\begin{abstract}
We consider channel estimation (CE) in narrowband Internet-of-Things (NB-IoT) systems. Due to the fluctuations in phase within receiver and transmitter oscillators, and also the residual frequency offset (FO) caused by discontinuous receiving of repetition coded transmit data-blocks, random phase noises are presented in received signals. Although the coherent-time of fading channel can be assumed fairly long due to the low-mobility of NB-IoT user-equipments (UEs), such phase noises have to be considered before combining the the channel estimates over repetition copies to improve their accuracies. In this paper, we derive a sequential minimum-mean-square-error (MMSE) channel estimator in the presence of random phase noise that refines the CE sequentially with each received repetition copy, which has a low-complexity and a small data storage. Further, we show through simulations that, the proposed sequential MMSE estimator improves the mean-square-error (MSE) of CE by 1 dB in the low signal-to-noise ratio (SNR) regime, compared to a traditional sequential MMSE estimator that does not thoroughly consider the impact of random phase noises.
\end{abstract}

\section{Introduction}

Narrowband Internet-of-Things (NB-IoT) is a new radio-access technology designed for ultra-low-end IoT applications \cite{Wetal17, 45820, 36201,36212}. In the latest Long-Term Evolution (LTE) Rel-13, the 3rd Generation Partnership Project (3GPP) \cite{36201} has introduced a number of key features for NB-IoT. A user-equipment (UE) operates in the downlink (DL) of NB-IoT system using 12 subcarriers with a subcarrier bandwidth 15 kHz, and in the uplink (UL) using a single subcarrier with bandwidth either 3.75 or 15 kHz, or alternatively 3, 6 or 12 subcarriers with a subcarrier bandwidth 15 kHz. A legacy global system for mobile communications (GSM) \cite{HR161} operator can have a \lq\lq{}stand-alone\rq\rq{} NB-IoT deployment by replacing one GSM carrier (200 kHz) with NB-IoT, while an LTE operator can allocate one physical resource blocks (PRBs) of 180 kHz to NB-IoT inside an LTE carrier, or deploying NB-IoT in the guard-band, which are know as \lq\lq{}inband\rq\rq{} and \lq\lq{}guard-band\rq\rq{} deployments, respectively. NB-IoT reuses the LTE design extensively to achieve a harmonious co-existence of current networks.

To offer a wide-area coverage and improve the robustness of data-transmissions, repetition coding \cite{36212} is used for both DL and UL in NB-IoT systems. The UE has to start decoding attempts for multiple possible repetitions which are limited by a maximum number of repetitions, which is 2048 for NB physical downlink shared channel (NPDSCH) and 128 for NB physical uplink shared channel (NPUSCH), respectively. There are three DL control channel indicator (CCI) formats specified which have different bit positions for the DCI information \cite{36212}, and the number of repetitions is written into a field of the DCI. The UE can therefore combine channel estimates over received repetition copies in the channel coherent-time to improve the CE accuracies and the data-transmission performance.

Due to the fluctuations in phase within receiver and transmitter oscillators and residual frequency offset (FO), there are random phase noises presented in received signals, which can severely degrade the mean-square-error (MSE) of channel estimation (CE). The performance losses in relation to signal-to-noise ratio (SNR) and bit error rate (BER) in OFDM systems have been well-studied in e.g., \cite{PM95, L98, TH01, SB04, GC98, PF04}. With NB-IoT systems considered, the problem is somewhat different from what has been considered before. Firstly, as NB-IoT systems only use 12 subcarriers for data-transmission, the phase noises can be approximately assumed to be identical through one PRB and yields a common phase error (CPE) estimation problem \cite{PF04}. Secondly, for low-end NB-IoT UEs, the CE algorithm needs to have a low computational cost. For instance, the maximum a \textit{posteriori} CE algorithm in \cite{LD06} and a majorization-minimization based joint phase noise and CE algorithm in \cite{WP17} are over complex and consumes too much power. Thirdly, the data-storage of the CE algorithm needs to be kept small in order to save the memory.

In this work, we consider CE in the presence of phase noise with repetition coded\footnote{Note that, as we only consider the reference symbols based CE, the proposed estimator also works for different transmit data-blocks as long as the channels are in the coherent-time.} transmit data-blocks in NB-IoT systems, and derive the sequential minimum-mean-square-error (MMSE) based CE using NB reference symbols (NRSs). The proposed sequential MMSE estimator has a low-complexity and only the correlation matrix needs to be stored and updated over each iteration. Further, the proposed estimator can be reviewed as an extension of a traditional sequential MMSE estimator with taking the impact of phase noise into consideration, and has a structure of a Kalman-filter \cite{K93}. Numerical results show that in the low the SNR regime, the proposed estimator performs around 1 dB better than a traditional sequential MMSE estimator that compensates the phase noise, but do not take their estimation accuracies into consideration.

\section{Received Signal Model in NB-IoT Systems}

\begin{figure}[t]
\begin{center}
\vspace*{-0.5mm}
\hspace*{12mm}
\scalebox{1.2}{\includegraphics{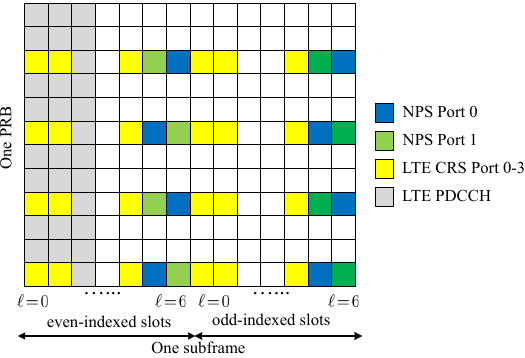}}
\vspace*{-2mm}
\caption{\label{fig1}The NRS pattern in one PRB (180 KHz) for stand-alone, guard-band or in-band deployments of NB-IoT systems in LTE Rel. 13.}
\vspace*{-7mm}
\end{center}
\end{figure}

We consider an NB-IoT system with a single transmit and receive antennas, and the NRS pattern is depicted in Fig. \ref{fig1}. For stand-alone and guard-band deployments, no LTE resource needs to be protected, and NPDCCH, NPDSCH or NRS can utilize all the resource elements (REs) in one PRB pair. However, for in-band deployment, NPDCCH, NPDSCH or NRS cannot be mapped to the REs taken by LTE cell-specific reference symbols (CRS) and control channels. NB-IoT is designed to allow a UE to learn the deployment mode and cell identity through initial acquisition.  

With phase noise $\phi_\ell[n]$ caused by oscillator fluctuations and a residual FO $f_e$ normalized by the subcarrier frequency, the time-domain baseband received signal at the $n$th sampling time of the $\ell$th orthogonal-frequency-division-multiplexing (OFDM) symbol can be expressed as
 \be  \label{sl} \tilde{s}_{\ell}[n]=e^{j\phi_\ell[n]}\!\!\left(\frac{1}{\sqrt{N}}\!\!\sum_{k=-N/2}^{N/2-1}\!S_{\ell}[k]e^{j\frac{2\pi n(f_e+k)}{N}}\right)\!\star h_\ell[n]+w(n), \ee
where \lq\lq{}$\star$\rq\rq{} denotes the linear convolution, $S_{\ell}[k]$ is the transmit symbol on the $\ell$th OFDM symbol and the $k$th subcarrier, $h[n]$ is discrete fading channel taps, $w(n)$ is additive-white-Gaussian-noise (AWGN), and $N$ is the Fast-Fourier-Transform (FFT) size. An FFT operation is implemented at the UE to obtain the received frequency domain signals,
{\setlength\arraycolsep{0pt}  \bea  \label{Sl}\tilde{S}_{\ell}[k]&=& \frac{1}{\sqrt{N}}\!\!\sum_{n=-\frac{N}{2}}^{\frac{N}{2}-1}\!\tilde{s}_{\ell}[n]e^{-\frac{j2\pi nk}{N}} \notag \\ &=&\frac{H_\ell[k]}{N}\!\!\sum_{n=-\frac{N}{2}}^{\frac{N}{2}-1}\sum_{m=-\frac{N}{2}}^{\frac{N}{2}-1}\!S_{\ell}[m]e^{j\phi_\ell[n]}e^{\frac{j2\pi n(f_e+m-k)}{N}}\!+\!\tilde{w}[k]\notag \\
&=&H_\ell[k]S_{\ell}[k]I_\ell(0)+\!\!\underbrace{\sum_{m=-\frac{N}{2},m\neq k}^{\frac{N}{2}-1}\!\!\!\!S_{\ell}[m]I_\ell[m]\!+\!\tilde{w}[k]}_{v[k]},
  \eea}
\hspace{-1.2mm}where $H_\ell[k]$ is the fading channels on frequency domain, $I_\ell[k]$ is a function of phase noises and FO, $v[k]$ comprises both the interference and noise. For each received repetition copy, although the channel taps $H_\ell[k]$ can be assumed to be identical as they are in the coherent-time, however, the term
\bea I_\ell(0)= \sum_{n=-\frac{N}{2}}^{\frac{N}{2}-1}e^{j\left(\phi_\ell[n]+\frac{2\pi n f_e}{N}\right)}\eea
changes due to the randomness of $\phi_\ell[n]$. As in NB-IoT, only the middle 12 subcarriers are used for data-transmission, $I_\ell(0)$ can be assumed to be identical for one received repetition copy. Further, as from \cite{PF04, SB04}, the module of $I_\ell(0)$ is very close to one, and can be seen as a symbol rotation in the complex plane. That is, we can approximate $I_\ell(0)$ by some $e^{j\phi_\ell}$. 

From (\ref{Sl}), stacking the received signal corresponding to all NPS symbols for a $m$th repetition copy in a vector form yields
 \bea  \label{md0} \tilde{\vec{S}}_m\approx e^{j\phi_m}\vec{h}\odot\vec{S}_m+\vec{v}_m, \eea
where \lq\lq{}$\odot$\rq\rq{} is the Hadamard product, $\tilde{\vec{S}}_m$ is the received signal vector, $\vec{S}_m$ is the transmitted NPS vector at the $m$th repetition copy with zero-means and satisfying $\mathbb{E}[\vec{S}_m\vec{S}_m\rmh]\!=\!\vec{I}$. The elements in the channel vector $\vec{h}$ are independent complex Gaussian variables with zero-means and unit-variances which can be assumed to be the same for a whole repetition period, the term $\vec{v}_m$ comprising both the interference and noise is modeled as AWGN with a variance $\gamma$ for all copies. The random phase rotation $\phi_m$ is a real value, whose probability density function (PDF), for analytical tractability, can be assumed to a uniform distribution over $[0,\,2\pi)$. For generality, we assume that there are $K$ NRS in one repetition copy.

By removing $\vec{S}_m$ from both sides in (\ref{md0}), we obtain a signal model corresponding to the least-square (LS) estimates of the channel vector as
 \bea  \label{md1} \vec{r}_m\approx e^{j\phi_m}\vec{h}+\tilde{\vec{v}}_m, \eea
where the noise $\tilde{\vec{v}}_m$ has the same statistical properties as $\vec{v}_m$. Given the received signal mode (\ref{md1}), we consider estimating $\vec{h}$ in a sequential manner based on each received repetition copy, which saves the required storage and also simplifies the operations for low-end NB-IoT UEs. 

\section{The Proposed Sequential MMSE Based CE}
Without loss of generality, we let $\phi_0\!=\!0$ for the first received copy, and denote $\hat{\vec{h}}$ as the CE obtained in a previous step. With a newly received $m$th copy, the vector $e^{j\phi_m}\vec{h}$ can be estimated. However, this estimate cannot be directly combined with $\hat{\vec{h}}$ due to the phase rotation $e^{j\phi_m}$. An intuitive way is to use $\hat{\vec{h}}$ to estimate $\phi_m$ first, and then obtain an update estimate $\hat{\vec{h}}$ after compensating the phase rotation. This can be implemented by a Kalman-filter based approach, but the optimal filtering needs to be solved. Instead, we directly derive a sequential MMSE estimator for $\vec{h}$ in he following.

The sequential MMSE estimator has the form \cite{K93} 
{\setlength\arraycolsep{2pt} \bea  \label{eq1} E(\vec{h}|\vec{r})&=&\int_{\mathbb{C}} \vec{h}p(\vec{h}|\vec{r}) \mathrm{d}\vec{h}\notag \\
&=&\int_{\mathbb{C}}\int_{\mathbb{R}}\frac{ \vec{h}p(\vec{r}|\vec{h},\phi)  p(\vec{h})p(\phi)}{p(\vec{r})}\mathrm{d}\vec{h}\mathrm{d}\phi\notag \\
&=&\frac{1}{p(\vec{r})}\int_{\mathbb{C}}\int_{\mathbb{R}}\vec{h}p(\vec{r}|\vec{h},\phi)  p(\vec{h})p(\phi)\mathrm{d}\vec{h}\mathrm{d}\phi,
\eea}
\hspace{-1.4mm}where the conditional PDF $p(\vec{r}|\vec{h},\phi)$ with a given pair ($\vec{h},\,\phi$) and under SNR $1/\gamma$ is complex Gaussian and equal to
\bea \label{pr} p(\vec{r}|\vec{h},\phi)=\frac{1}{(\pi \gamma)^K}\exp\left( -\frac{\|\vec{r}-\vec{h}\exp(j\phi)\|^2}{\gamma}\right).\eea
Assuming that $\hat{\vec{h}}$ is the estimate of $\vec{h}$ obtained with the previous received copy, the PDF $p(\vec{h})$ is also complex Gaussian and can be estimated by
\bea \label{ph} p(\vec{h}|\hat{\vec{h}})=\frac{1}{(\pi)^K\det(\vec{R})}\exp\left( -(\hat{\vec{h}}-\vec{h})\rmh\vec{R}^{-1}(\hat{\vec{h}}-\vec{h})\right), \eea
where $\vec{R}\!=\!\mathbb{E}[\vec{h}\vec{h}\rmh]$ denotes the correlation matrix of $\vec{h}$.

\begin{figure*}[b]
\vspace{-4mm}
\hrulefill
{\setlength\arraycolsep{0pt} \bea \label{int1} &&\int_{\mathbb{C}}\int_{\mathbb{R}} \vec{h}p(\vec{r}|\vec{h},\phi)  p(\vec{h})p(\phi)\mathrm{d}\vec{h}\mathrm{d}\phi\notag \\
&&=\frac{1}{(\pi \gamma)^K}\frac{1}{(\pi)^K\det(\vec{R})}\int_{\mathbb{C}}\int_{\mathbb{R}} \vec{h}\exp\left( -\frac{\|\vec{r}-\vec{h}\exp(j\phi)\|^2}{\gamma}\right) \exp\left( -(\hat{\vec{h}}-\vec{h})\rmh\vec{R}^{-1}(\hat{\vec{h}}-\vec{h})\right) \mathrm{d}\vec{h}\mathrm{d}\phi  \notag \\
&&=\frac{1}{(\pi \gamma)^K}\frac{1}{(\pi)^K\det(\vec{R})}\exp\left(-\frac{1}{\gamma}\|\vec{r}\|^2-\hat{\vec{h}}\rmh\vec{R}^{-1}\hat{\vec{h}}\right) \notag \\
&& \qquad\qquad\qquad\qquad \times
\underbrace{\int_{\mathbb{C}}\int_{\mathbb{R}} \vec{h}\exp\left(-\vec{h}\rmh\left(\vec{R}^{-1}+\frac{1}{\gamma}\vec{I}\right)\vec{h}+2\Re\left\{\left(\frac{1}{\gamma}\vec{r}\rmh\exp(j\phi)+\hat{\vec{h}}\rmh\vec{R}^{-1}\right)\vec{h}  \right\} \right)p(\phi)  \mathrm{d}\vec{h}\mathrm{d}\phi}_{\xi_1}  \eea}
\hrulefill
{\setlength\arraycolsep{0pt} \bea \label{int2} p(\vec{r})&=&\int_{\mathbb{C}}\int_{\mathbb{R}} p(\vec{r}|\vec{h},\phi)  p(\vec{h})p(\phi)\mathrm{d}\vec{h}\mathrm{d}\phi\notag \\
&&=\frac{1}{(\pi \gamma)^K}\frac{1}{(\pi)^K\det(\vec{R})}\int_{\mathbb{C}}\int_{\mathbb{R}} \exp\left( -\frac{\|\vec{r}-\vec{h}\exp(j\phi)\|^2}{\gamma}\right) \exp\left( -(\hat{\vec{h}}-\vec{h})\rmh\vec{R}^{-1}(\hat{\vec{h}}-\vec{h})\right) \mathrm{d}\vec{h}\mathrm{d}\phi  \notag \\
&&=\frac{1}{(\pi \gamma)^K}\frac{1}{(\pi)^K\det(\vec{R})}\exp\left(-\frac{1}{\gamma}\|\vec{r}\|^2-\hat{\vec{h}}\rmh\vec{R}^{-1}\hat{\vec{h}}\right) \notag \\
&& \qquad\qquad\qquad\qquad \times
\underbrace{\int_{\mathbb{C}}\int_{\mathbb{R}} \exp\left(-\vec{h}\rmh\left(\vec{R}^{-1}+\frac{1}{\gamma}\vec{I}\right)\vec{h}+2\Re\left\{\left(\frac{1}{\gamma}\vec{r}\rmh\exp(j\phi)+\hat{\vec{h}}\rmh\vec{R}^{-1}\right)\vec{h}  \right\} \right)p(\phi)  \mathrm{d}\vec{h}\mathrm{d}\phi}_{\xi_2}  \eea}
\vspace{-12mm}
\end{figure*}

With (\ref{pr}) and (\ref{ph}), the integrals in (\ref{eq1}) and $p(\vec{r})$ are expressed in (\ref{int1}) and (\ref{int2}), respectively. Therefore, the estimator (\ref{eq1}) is
\bea  \label{eq2} E(\vec{h}|\vec{r})=\frac{\xi_1}{\xi_2},\eea
where $\xi_1$ and $\xi_2$ are further derived in (\ref{int3}) and (\ref{int4}), respectively. Inserting them back into (\ref{eq2}) yields a final form of the proposed MMSE estimator, which is stated in Theorem 1.
\begin{figure*}[b]
\vspace{-4mm}
\hrulefill
{\setlength\arraycolsep{0pt} \bea  \label{int3}
\xi_1&=&\frac{ \left(\vec{R}^{-1}+\frac{1}{\gamma}\vec{I}\right)^{-1}}{\det\left(\vec{R}^{-1}+\frac{1}{\gamma}\vec{I}\right)}\int_{\mathbb{R}}\left(\frac{1}{\gamma}\vec{r}\exp(-j\phi)+\vec{R}^{-1}\hat{\vec{h}}\right)\notag \\
&& \qquad\qquad\qquad\qquad \times \exp\left(\left(\frac{1}{\gamma}\vec{r}\rmh\exp(j\phi)+\hat{\vec{h}}\rmh\vec{R}^{-1}\right)\left(\vec{R}^{-1}+\frac{1}{\gamma}\vec{I}\right)^{-1}\left(\frac{1}{\gamma}\vec{r}\exp(-j\phi)+\vec{R}^{-1}\hat{\vec{h}}\right)\right)p(\phi)\mathrm{d}\phi   \notag \\
&=&\frac{ \left(\vec{R}^{-1}+\frac{1}{\gamma}\vec{I}\right)^{-1}}{\det\left(\vec{R}^{-1}+\frac{1}{\gamma}\vec{I}\right)}\exp\left(\frac{1}{\gamma}\vec{r}\rmh\left(\gamma\vec{R}^{-1}+\vec{I}\right)^{-1}\vec{r}\right) \exp\left(\hat{\vec{h}}\rmh\left(\vec{I}+\frac{1}{\gamma}\vec{R}\right)^{-1}\vec{R}^{-1}\hat{\vec{h}}\right) \notag \\
&& \qquad \times \int_{\mathbb{R}}\left(\frac{1}{\gamma}\vec{r}\exp(-j\phi)+\vec{R}^{-1}\hat{\vec{h}}\right) \exp\left(2\Re\left\{\frac{\exp(j\phi)}{\gamma}\vec{r}\rmh\left(\vec{I}+\frac{\vec{R}}{\gamma}\right)^{-1}\hat{\vec{h}}\right\}\right)p(\phi)\mathrm{d}\phi.\eea}
\hrulefill
{\setlength\arraycolsep{0pt} \bea \label{int4}
\xi_2&=&\frac{1}{\det\left(\vec{R}^{-1}+\frac{1}{\gamma}\vec{I}\right)}\int_{\mathbb{R}} \exp\left(\left(\frac{1}{\gamma}\vec{r}\rmh\exp(j\phi)+\hat{\vec{h}}\rmh\vec{R}^{-1}\right)\left(\vec{R}^{-1}+\frac{1}{\gamma}\vec{I}\right)^{-1}\left(\frac{1}{\gamma}\vec{r}\exp(-j\phi)+\vec{R}^{-1}\hat{\vec{h}}\right)\right)\mathrm{d}\phi   \notag \\
&=&\frac{1}{\det\left(\vec{R}^{-1}+\frac{1}{\gamma}\vec{I}\right)}\exp\left(\frac{1}{\gamma}\vec{r}\rmh\left(\gamma\vec{R}^{-1}+\vec{I}\right)^{-1}\vec{r}\right) \exp\left(\hat{\vec{h}}\rmh\left(\vec{I}+\frac{1}{\gamma}\vec{R}\right)^{-1}\vec{R}^{-1}\hat{\vec{h}}\right) \notag \\
&& \qquad \times \int_{\mathbb{R}}\exp\left(2\Re\left\{\frac{\exp(j\phi)}{\gamma}\vec{r}\rmh\left(\vec{I}+\frac{\vec{R}}{\gamma}\right)^{-1}\hat{\vec{h}}\right\}\right)p(\phi)\mathrm{d}\phi.
\eea}
\vspace{-12mm}
\end{figure*}
\begin{theorem}
The sequential MMSE estimator of channel vector $\vec{h}$ in the presence of random phase noise is
\bea \label{Ehr} E(\vec{h}|\vec{r})=\tilde{\vec{R}}\left(\hat{\vec{h}}+\frac{\zeta}{\gamma}\vec{R}\vec{r} \right),\eea
where $\hat{\vec{h}}$ is the estimate of $\vec{h}$ obtained in the previous step with $\vec{R}$ being its correlation matrix, $1/\gamma$ is the SNR, and $\vec{r}$ is a newly received repetition copy. The MSE matrix
\bea \tilde{\vec{R}}=\left(\vec{I}+\frac{\vec{R}}{\gamma}\right)^{-1},\eea
and the factor $\zeta$ equals
\bea  \label{zeta} \zeta=\frac{\beta_1}{\beta_2}\frac{\vec{r}\rmh\tilde{\vec{R}}\hat{\vec{h}}}{\left|\vec{r}\rmh\tilde{\vec{R}}\hat{\vec{h}}\right|},\eea
with $\beta_1$ and $\beta_2$ calculated as
{\setlength\arraycolsep{2pt} \bea \beta_1&=&\int_{\mathbb{R}}\!\exp\!\left(\!-j\phi\!+\!2\Re\left\{\frac{\exp(j\phi)}{\gamma}\vec{r}\rmh \tilde{\vec{R}} \hat{\vec{h}}\right\}\!\right)\!p(\phi)\mathrm{d}\phi, \qquad
\\
\beta_2&=&\int_{\mathbb{R}}\!\exp\!\left(2\Re\left\{\frac{\exp(j\phi)}{\gamma}\vec{r}\rmh\tilde{\vec{R}}\hat{\vec{h}}\right\}\!\right)p(\phi)\mathrm{d}\phi.\eea}
\end{theorem}

To obtain closed-from expressions for $\beta_1$ and $\beta_2$, we assume $\phi$ to be uniformly distributed in $[0,\,2\pi)$ which yields
{\setlength\arraycolsep{2pt} \bea \beta_1&=&I_1\!\left(\frac{2}{\gamma}\left|\vec{r}\rmh\tilde{\vec{R}}\hat{\vec{h}}\right|\!\right),
\\
\beta_2&=&I_0\!\left(\frac{2}{\gamma}\left|\vec{r}\rmh\tilde{\vec{R}}\hat{\vec{h}}\right|\!\right),\eea
using \cite[Formulas 3.937.1 and 3.937.2]{AS72}, where $I_1$ and $I_0$ are the modified Bessel functions of the first kind. 

The variable $\zeta$ in (\ref{zeta}) comprises two parts: the first part is to estimate the phase noise with $\vec{r}\rmh\tilde{\vec{R}}\hat{\vec{h}}\big/|\vec{r}\rmh\tilde{\vec{R}}\hat{\vec{h}}|$, and the second part is to scale down the estimate by a factor $I_1\!\left(\frac{2}{\gamma}\left|\vec{r}\rmh\tilde{\vec{R}}\hat{\vec{h}}\right|\right)\!\Big/\!I_0\!\left(\frac{2}{\gamma}\left|\vec{r}\rmh\tilde{\vec{R}}\hat{\vec{h}}\right|\right)$, which is smaller than 1 and asymptotically equals 1 as SNR increases.

Now let\rq{}s consider two simple cases to illustrate the sequential MMSE estimator (\ref{Ehr}). The first case is $\vec{R}\!=\!\vec{I}$, that is, the elements in $\vec{h}$ are identical and independently distributed (IID). In this case, (\ref{Ehr}) becomes
 {\setlength\arraycolsep{2pt}\bea\label{Ehr1}  E(\vec{h}|\vec{r})
&=&\left(1+\frac{1}{\gamma}\right)^{-1}\left(\hat{\vec{h}}+\frac{\zeta_1}{\gamma}\vec{r} \right) \notag \\
&=&\frac{\gamma}{\gamma+1}\hat{\vec{h}}+\frac{\zeta_1}{\gamma+1}\vec{r},
\eea}
where
\bea  \zeta_1=\frac{I_1\!\left(\frac{2\left|\vec{r}\rmh\hat{\vec{h}}\right|}{\gamma+1}\right)}{I_0\!\left(\frac{2\left|\vec{r}\rmh\hat{\vec{h}}\right|}{\gamma+1}\right)}\frac{\vec{r}\rmh\hat{\vec{h}}}{\left|\vec{r}\rmh\hat{\vec{h}}\right|}.
\eea
The estimator (\ref{Ehr1}) is the traditional sequential MMSE estimator \cite{K93}, but with an extra factor $\zeta_1$ which is introduced by the random phase.

The second case is $\vec{R}\!=\!\vec{1}_{K\times K}$, that is, the elements in $\vec{h}$ are fully dependent and only one scalar $h$ needs to be estimated. Following the same derivations leading to (\ref{Ehr}), it can be shown that the sequential MMSE estimator in this case is
\bea E(h|\vec{r})=\frac{\gamma}{\gamma+1}\hat{h}+\frac{\zeta_2}{\gamma+1}\tilde{r}, \eea
where $\tilde{r}$ is the mean value of $\vec{r}$ and
\bea  \zeta_2=\frac{I_1\left(\frac{2K\left|\tilde{r}\hat{h}\right|}{\gamma+1}\right)}{I_0\left(\frac{2K\left|\tilde{r}\hat{h}\right|}{\gamma+1}\right)}\frac{\tilde{r}\rmh\hat{h}}{\left|\tilde{r}\rmh\tilde{h}\right|}.
\eea
This is equivalent to the first case with a single observation $\tilde{r}$, but both the variances of $h$ and the noise are decreased by a factor of $K$, which make an intuitive sense.

Using Theorem 1, after $E(\vec{h}|\vec{r})$ is computed, the correlation matrix $\vec{R}$ needs to be updated at each step. By assuming that $\zeta$ results in a perfect estimation of the phase and as $\beta_1/\beta_2\!\approx\!1$ at high SNR, $\vec{R}$ can be updated according to
\bea \vec{R}_{m+1}=\vec{R}_m\!\left(\vec{I}+\frac{\vec{R}_m}{\gamma}\right)^{-1}\!.\eea
The initialization of $\vec{R}$ can be set to a heuristic value or estimated through a long-term averaging. With $\hat{\vec{h}}_m$, the estimate of phase noise $\phi_m$ of the $m$th repetition copy is obtained as
\bea \hat{\phi}_{m}=\angle\!\left(\hat{\vec{h}}_{m}\rmh\vec{r}_m\right)\!,\eea
where \lq\lq{}$\angle$\rq\rq{} takes the angle of the input variable. 

\vspace{-1.5mm}
\begin{algorithm}[ht!]
	\caption{The proposed sequential MMSE estimator.}
	\label{alg:2}
      \begin{algorithmic}[1]
      \STATE Initialize $m\!=\!0$ and $\hat{\vec{h}}_m\!=\!\vec{0}$, and assume the channel correlation matrix $R_0$ and $\gamma$ are known. \\
      \STATE With the $(m\!+\!1)$th received repetition copy $\vec{r}_{m+1}$, update the parameters according to
       {\setlength\arraycolsep{2pt}\bea
       \tilde{\vec{R}}_{m+1}&=&\left(\vec{I}+\frac{\vec{R}_m}{\gamma}\right)^{-1},\notag \\
       \hat{\vec{h}}_{m+1}
&=& \tilde{\vec{R}}_{m+1}\left( \hat{\vec{h}}_{m}+\frac{ \zeta}{\gamma}\vec{R}_{m}\vec{r}_{m+1} \right)\!, \notag \\
       \vec{R}_{m+1}&=&\vec{R}_m\!\left(\vec{I}+\frac{\vec{R}_m}{\gamma}\right)^{-1}. \notag \eea}
       where $\zeta$ is calculated according to (\ref{zeta}).
       \STATE Repeat Step 2 until the transmit data-block is successively decoded or all repetition codes have been received.
\end{algorithmic}
\vspace{-1mm}
\end{algorithm}
\vspace{-2mm}

Based on Theorem 1, the proposed sequential MMSE estimator is summarized in Algorithm 1. Note that, under the case that there is no phase noise $\phi$ presented, we can set $\zeta\!=\!1$ and then Algorithm 1 is identical to the traditional sequential MMSE estimator \cite{K93} for $\vec{h}$.

\section{Numerical Results}
In this section, we evaluate MSE of the proposed sequential MMSE estimator in comparison to the traditional sequential MMSE estimator, which refers to an estimator that only compensates the phase noise by setting $\zeta$ to
\bea  \label{zeta1} \zeta=\frac{\vec{r}\rmh\tilde{\vec{R}}\hat{\vec{h}}}{\left|\vec{r}\rmh\tilde{\vec{R}}\hat{\vec{h}}\right|}\!,\eea
in Algorithm 1, that is, setting $\beta_1/\beta_2\!=\!1$ in (\ref{zeta}) and ignores the impact of $\beta_1$ and $\beta_2$, i.e., the quality of the phase estimates. In addition, we also evaluate the MSE under ideal cases when there is no phase noise presented, in which cases, the proposed sequential MMSE and traditional estimators are identical. The MSE at the $m$th repetition copy is measured\footnote{The MSE can also be equivalently calculated as the trace of $\vec{R}_{m}$.} according to
\be \mathrm{MSE}(m)\!=\!\frac{1}{NK}\!\sum_{n=0}^{N-1}\!\!\left(\!\left\|\hat{\vec{h}}_{n,m}\!\exp(j\hat{\phi}_{n,m})\!-\!\vec{h_n}\!\exp(j\phi_{n,m})\right\|^2\!\right)\!, \notag \ee
where we simulate $N\!=\!20000$ channel realizations, and $\hat{\vec{h}}_{n,m}$, $\hat{\phi}_{n,m}$ are estimates of $\vec{h}_n$ and $\phi_{n,m}$ at the $m$th repetition copy.

In Fig. \ref{fig2}, we evaluate the MSE of the proposed and traditional sequential MMSE estimators, and the MSE obtained when there is no phase noise present. We test under AWGN channels with SNR equals to -4, -2 and 0 dB respectively. As can be seen, the proposed sequential MMSE estimator has around 1 dB gain over the traditional estimator without considering $\beta_1/\beta_2$ in (\ref{zeta}) at SNR -4 dB. In Fig. \ref{fig3}, we repeat the tests in Fig. \ref{fig2} under ETU-3Hz channels with SNR equal to -3, 0 and 3 dB, respectively. As can be seen, the proposed estimator also outperforms the traditional estimator around 1 dB at SNR -3 dB. 

In both Fig. \ref{fig2} and \ref{fig3}, when SNR increases, the proposed sequential MMSE estimator starts to perform close to the traditional estimator, which is due to the fact that $\beta_1/\beta_2$ quickly converges to 1 as SNR increases. Further, as the number of received repetition copies increases, the MSE renders error-floors under both cases compared to the ideal cases, as a consequence of the presence of phase noises. 

As shown in Fig. \ref{fig3}, the MSE converges within 10$\sim$20 repetition copies. In Fig. \ref{fig4}, we plot the MSE obtained at the 20th received copy under ETU-3Hz channels and at different SNR values. As can be seen, when SNR is less than 0 dB, the proposed sequential MMSE estimator renders considerable gains over the traditional estimator, with only a minor complexity increment to compute the values $\beta_1/\beta_2$ through e.g., look-up-table operations. We also investigate the impact of different initializations of $\vec{R}_0$. Setting $\vec{R}_0$ to the ideal correlation matrix yields around 2 dB gains over the case initializing $\vec{R}_0$ with an identity matrix. When there is no phase noise present, the MSE gains are larger with ideal $\vec{R}_0$.
 
\begin{figure}[t]
\begin{center}
\vspace*{-2mm}
\hspace*{-8mm}
\scalebox{0.328}{\includegraphics{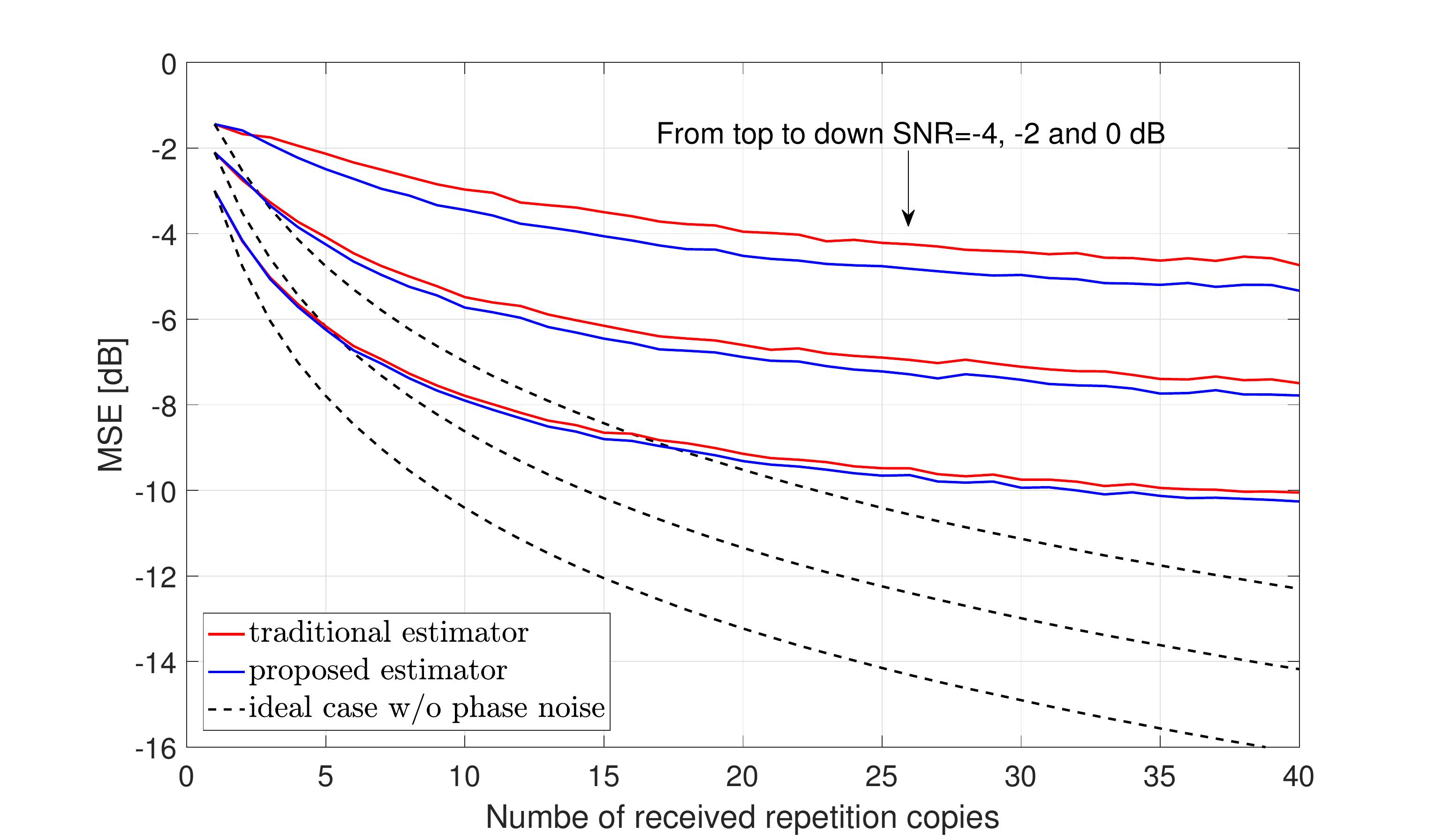}}
\vspace*{-7mm}
\caption{\label{fig2}The evaluation of MSE under AWGN channel.}
\vspace*{-7mm}
\end{center}
\end{figure}

\section{Summary}
We have considered channel estimation (CE) in narrowband Internet-of-Things (NB-IoT) systems in the presence of random phase noise caused by the fluctuations of oscillators and the residual frequency offset (FO). We have derived a sequential minimum-mean-square-error (MMSE) estimator for the CE, which updates the estimates in a sequential way with a low computational cost and a small data storage. Moreover, we show through simulations that, the proposed sequential MMSE estimator performs 1 dB better in the low signal-to-noise ratio (SNR) regime in terms of mean-square-error (MSE) than a traditional sequential MMSE estimator that did not thoroughly consider the impact of the phase noises.

\begin{figure}[t]
\begin{center}
\vspace*{-2mm}
\hspace*{-5mm}
\scalebox{0.328}{\includegraphics{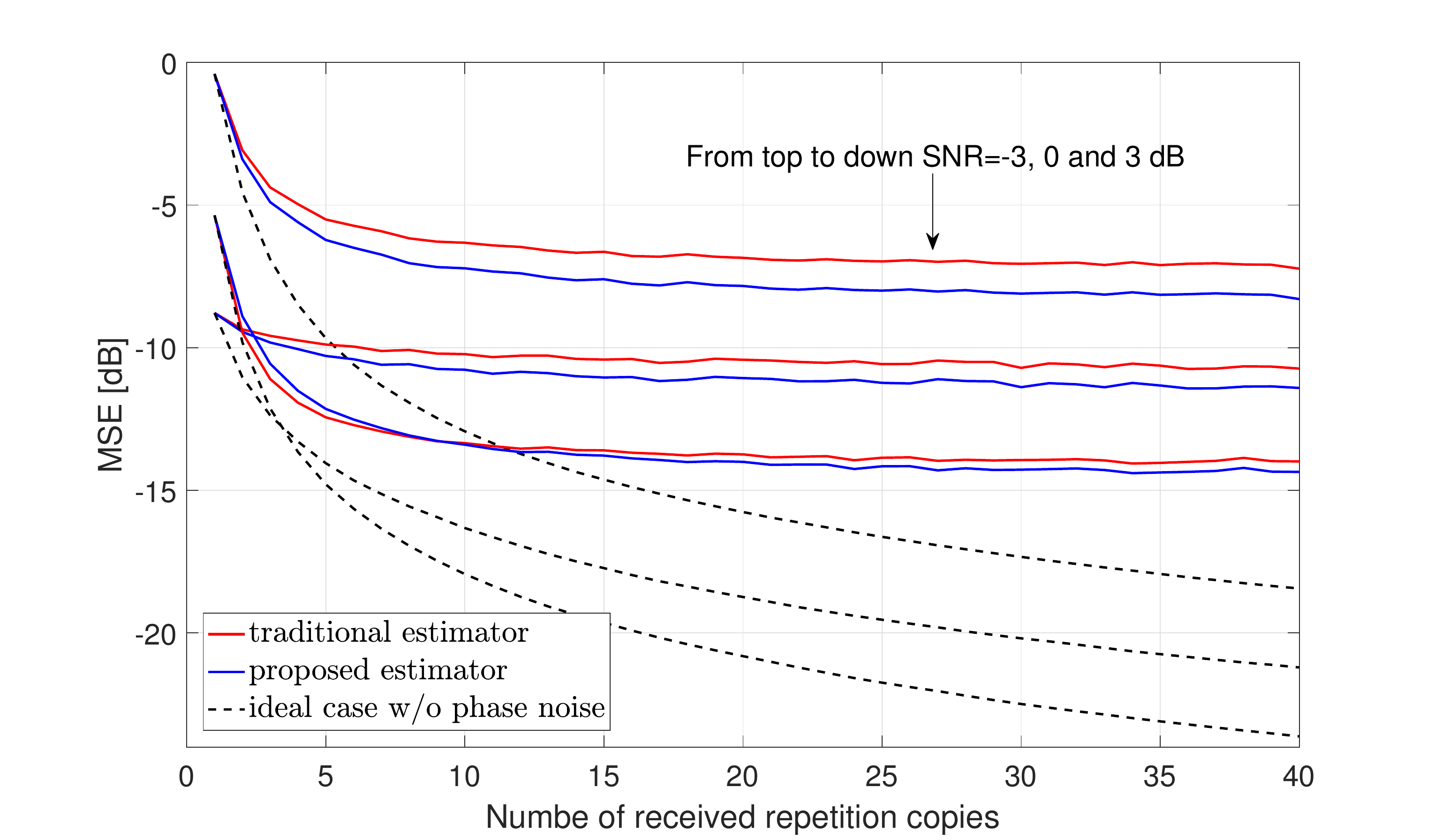}}
\vspace*{-7mm}
\caption{\label{fig3}The evaluation of MSE under ETU-3Hz channel.}
\vspace*{-4mm}
\end{center}
\end{figure}

\begin{figure}
\begin{center}
\vspace*{-2mm}
\hspace*{-5mm}
\scalebox{0.328}{\includegraphics{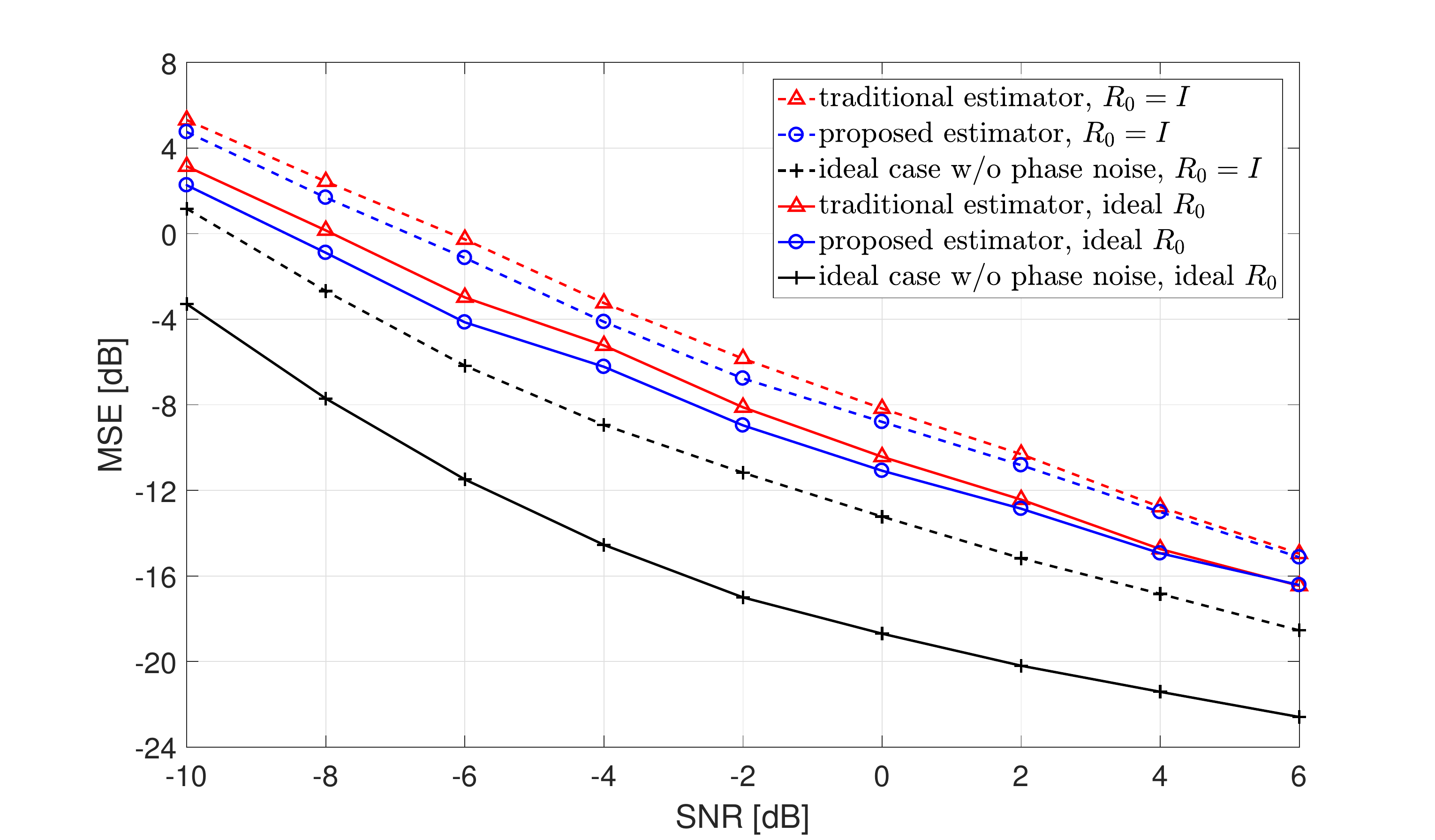}}
\vspace*{-7mm}
\caption{\label{fig4}The MSE measured at the 20th received copy under ETU-3Hz channel.}
\vspace*{-8mm}
\end{center}
\end{figure}

\end{document}